\documentclass[aip,pop,amsmath,amssymb,reprint]{revtex4-1}
\usepackage{graphicx}

\begin{document}

\title[Coulomb collisions in partially ionized gases]{First principles calculation of the effect of Coulomb collisions\\ in partially ionized gases}

\author{Z. Donk\'o}
\affiliation{Institute for Solid State Physics and Optics, Wigner Research Centre for Physics, Hungarian Academy of Sciences, P.O.B. 49, H-1525 Budapest, Hungary}

\date{\today}

\begin{abstract}
Coulomb collisions, at appreciable ratios ($\eta$) of the electron to the neutral particle density, influence significantly the electron kinetics in particle swarms and in plasmas of gas discharges. This paper introduces a combination of Molecular Dynamics and Monte Carlo simulation techniques, to provide a novel, approximation-free, first principles calculation method for the velocity distribution function of electrons, and related swarm characteristics, at arbitrary $\eta$. Simulation results are presented for electrons in argon gas, for density ratios between zero and 10$^{-1}$, representing the limits of a negligible electron density and an almost complete Maxwellization of the velocity distribution function, respectively.
\end{abstract}  



\maketitle

\section{Introduction}

Electron transport in gases, under the effect of electric and/or magnetic fields, has been attracting continuous interest since the early years of gas discharge physics. 

Exact  description of electron transport (including boundary effects, relaxation phenomena, non-equilibrium effects, etc.) requires a kinetic approach. \cite{noneq,Robson,Zoran09,Dujko11} The two powerful, equivalent, but quite different approaches to this problem are (i) calculations based on the Boltzmann equation (BE), and (ii) simulations based on Monte Carlo (MC) techniques. Both methods make it possible to obtain the central quantity of the kinetic theory, the {\it velocity distribution function}, $f({\bf v},{\bf r},t)$, of the particles. BE methods calculate $f$ directly, while MC methods follow a large number of individual particles to ``build up'' $f$ from sampling and averaging particle phase space coordinates. Both BE and MC methods have been used, separately or jointly, for the description of a wide variety of phenomena in ionized gases, and have also been cross checked with each other in a number of studies. \cite{Reid,Pinhao,Trunec}

The effects of {\it electron-electron collisions} on swarm and discharge plasma characteristics have been considered only in a fraction of studies. These works have concluded that Coulomb collisions can be neglected at low values of the electron density ($n_e$) to neutral particle number density ($n$) ratio, $\eta = n_e / n \lesssim 10^{-6}$. However, in settings characterized by a higher $\eta$, like the positive column and negative glow regions of DC gas discharges, bulk plasma regions of high-power radio-frequency discharges, as well as arc discharges, they may influence electron kinetics to a considerable extent. In the DC negative glow, e.g., the cold-electron temperature is established by a balance between cooling in elastic collisions with gas atoms, and heating due to Coulomb collisions with hot electrons. \cite{Lawler} Coulomb collisions, likewise, influence the trapping of electrons in potential wells related to striations in gas discharges, \cite{Kolobov,Golubovskii} modify the velocity distribution function and transport coefficients of electrons \cite{Loffhagen1}, and affect the development of negative differential conductivity. \cite{Aleksandrov} They play an import role as well in dense plasmas.\cite{dense1}
 
Inclusion of Coulomb collisions in either the BE, or the MC approaches represents a significant challange, due to the long-range nature of the Coulomb potential. One possible simplification is to treat the multiple electron-electron interactions as a succession of {\it discrete, binary} collisions; this approximation has often been adopted in Boltzmann equation analysis, \cite{Rockwood,Yousfi1,Loffhagen1,Hagelaar} as well in Monte Carlo simulations. \cite{Kushner, Hashiguchi} Different methods for a more efficient description of Coulomb collisions have also been proposed.\cite{Nanbu1,Caflish} 
 
This paper introduces a novel, {\it approximation-free method} for the description of the motion of electrons in a background gas, under the influence of a static external electric field and electron-electron interactions, at arbitrary ratios of the electron to the neutral particle densities. The method is based on a combination of the classical Monte Carlo technique and a Molecular Dynamics method, both of which are based on first principles, and have extensively been used in different areas of plasma physics. Section II describes the methods and their combination. The results are presented in Sec. III, while a brief summary is given in Sec. IV.

\section{Simulation method}

While MC simulations \cite{Longo} have been routinely applied for the description of electron swarms, Molecular Dynamics (MD) methods have not been used so far for such purpose, to the best knowledge of the author. MD methods\cite{md} are well suited for the simulation of many-particle systems. By following the time-dependent trajectories of the particles in the phase space, MD simulations can generate pair correlation functions, make it possible to obtain transport coefficients, and allow the identification of collective excitations and the calculation of their dispersion relations. \cite{sccs} Thus, the present combination of MC and MD approaches provides new insights into the physics of particle swarms via the calculation of quantities that have not been accessible from either BE or MC calculations. Our approach, in particular, allows observation of the (i) onset of pair correlations between electrons and (ii) emergence wave phenomena in electron swarms.  

The capabilities of the present method are illustrated on a model system of a swarm of electrons in argon gas, for a wide domain of the {\it density ratio} $\eta$ The classical electron gas present in our model system is described by an MD procedure, as a many-body system. Ions are not accounted for in these calculations, electrons interact via the (un-screened) Coulomb potential. The interaction of the electron gas with the buffer gas is handled by a Monte Carlo collision procedure that is well suited for the short-range interaction of electrons with the gas atoms. The model system is assumed to be homogeneous and infinite, in other words, we establish a zero-dimensional model, in which two parameters, the reduced electric field ($E/n$) and the ratio of the electron density to the neutral density ($\eta$) fully characterize the system. 

In the MD simulation the phase space trajectories of the (classical) electrons ($i = 1,\dots,N$) are followed via the integration of their equations of motion:
\begin{equation}
m \frac{{\rm d}^2 {\bf r}_i}{{\rm d}t^2} = \sum_{i \neq j} {\bf F}_{ij} + e {\bf E},
\label{eq:eom}
\end{equation}
where the sum gives the force acting on particle $i$ by all other particles, and $e$ is the (negative) charge of the electron. The electric field, ${\bf E} = - {\bf e}_x E$, is homogeneous, the ${\bf e}_x$ unit vector points into the $x$ direction. (Note that in the absence of the interaction of the electrons with a background gas the electrons would continuously be accelerated due to the external field, which will, however, not be the case when e$^-$+Ar atoms collisions take place.) The motion of the electrons is simulated inside a cubic box, with periodic boundary conditions. The determination of the long-ranged Coulomb forces acting on the particles is the key question in the method, as calculation of the interparticle forces requires summation not only over all other particles within the box, but also over all the periodic replicas of the simulation box, to infinity. 

We adopt the Particle-Particle Particle-Mesh (PPPM) approach \cite{HE} to solve this problem. The simulations describe a micro-canonical ensemble, where the number of particles, the volume of the system and energy are conserved. The absolute size of the simulation box does not play a role. The upper limit for the time step, $\Delta t$, of the integration of (\ref{eq:eom}) is set by the stability requirement, at the closest approach of two particles, $r_{min} = e^2/(4 \pi \epsilon_0 \varepsilon_{max})$. Here $\varepsilon_{max}$ is a pre-defined maximum energy \cite{HE}, which has to be chosen carefully, to ensure that the probability of finding electrons with $\varepsilon > \varepsilon_{max}$ is vanishingly small at the conditions considered.

The electron gas and the background gas interact via e$^-$+Ar collisions. We adopt the simplified cross section set \cite{Phelps} that includes elastic collisions, excitation to one effective level, and ionization. The probability of an e$^-$+Ar collision during a time step $\Delta t$ is calculated as:\cite{donko}
\begin{equation}
P_{\rm coll} = 1 - {\rm exp}[-n_g \sigma_t(v) v \Delta t],
\end{equation}
where $\sigma_t$ is the total cross section, and $v$ is the actual velocity of the given electron. $P_{\rm coll}$ is calculated for each electron in each time step, and decision about the occurrence of a collision is made by comparing it with a random number. The type of the collision is selected randomly, based on the magnitudes of the cross sections of the individual possible processes, at the given electron velocity. All types of collisions are assumed to scatter electrons isotropically. As a simplification, ionization, is treated as a conservative process, and we adopt the cold-gas approximation  in the numerical description of e$^-$+Ar collisions.

We cover the 5 Td$\leq E/n \leq$ 20 Td domain of the reduced electric field, relevant to swarm conditions and to low-field regions of gas discharges. The electron to neutral density ratio ($\eta$) is varied within the range from zero to 10$^{-1}$. We use $\varepsilon_{max}$ = 35 eV for the maximum electron energy, that defines the simulation time step $\Delta t$. (The correctness of this choice is confirmed by the simulation results.) We follow the motion of $N$ = 10,000 electrons inside a cubic simulation box. (Using of a fixed number of particles is made possible by treating ionization as a conservative process.) The temperature of the background gas is $T$ = 300 K.

\section{Results}

The simulation runs consist of an (i) initial equilibration phase and a subsequent (ii) measurement phase. At the initialization of the simulations each electron is placed at a random position inside the simulation box, and is assigned to have $\varepsilon_{init}$ = 1 eV energy and a velocity vector pointing in a random direction over a unit sphere. As this is clearly far from the equilibrium velocity distribution, the system needs time to equilibrate. This equilibration (as well as the stability of the simulation) is monitored by calculating the time-dependence of the first four moments of the instantaneous $f(v,t)$ distribution function, $\langle v^k \rangle$. These moments are normalized by those characterizing a Maxwellian distribution: 
\begin{eqnarray}
\langle v \rangle_M  = 2 \alpha \sqrt{2/\pi}, ~~&~~ \langle v^2 \rangle_M  = 3 \alpha^2, \\ \nonumber
\langle v^3 \rangle_M  = 8 \alpha^3 \sqrt{2/\pi}, ~~&~~ \langle v^4 \rangle_M  = 15 \alpha^4,
\end{eqnarray} 
where $\alpha = \sqrt{2\langle \varepsilon\rangle / 3 m_e}$, $\langle \varepsilon\rangle$ is the mean electron energy, and $m_e$ is the electron mass. The normalized moments convey information about the ``similarity'' of the distribution function $f(v)$ obtained for the given conditions, with a Maxwellian. In the calculation of the velocity moments the ``instantaneous drift velocity'' (average $v_x$ at a given time) is subtracted from the velocities of the individual particles (thus for a drifting Maxwellian distribution all normalized moments are equal to 1).

\begin{figure}[h!]
\includegraphics[width=1\columnwidth]{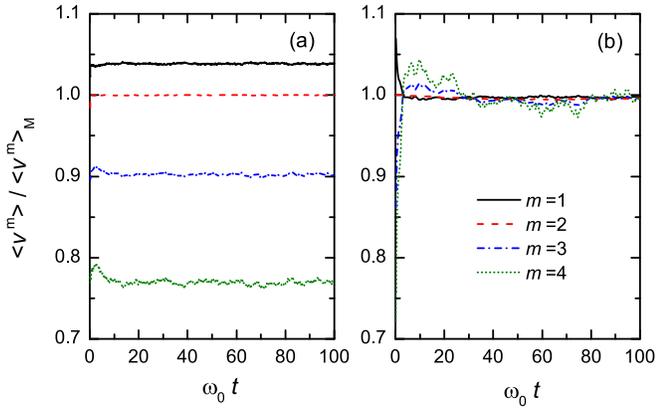}
\caption{\label{fig:relax}
(color online) Normalized velocity moments of $f(v,t)$: (a) $\eta = 10^{-6}$, (b) $\eta = 10^{-1}$. $E/n$ = 10 Td.}
\end{figure}

The equilibration of the swarm is illustrated in Fig.~\ref{fig:relax}, for $E/n$=10 Td and $\eta = 10^{-6}$ [panel (a)] and $\eta = 10^{-1}$ [panel (b)], time is normalized by the (electron) plasma frequency 
\begin{equation}
\omega_0=\sqrt{n_e e^2 / \epsilon_0 m_e}.
\end{equation}
The relaxation of the system is quite fast, and the stability of the simulation is confirmed in both cases. At $\eta = 10^{-6}$ the shape of $f(v)$, mostly determined by e$^{-}$--Ar collisions, stabilizes far from Maxwellian, as indicated by a large deviation of the velocity moments from those characterizing a Maxwellian distribution. (The stable value of the second moment, $\langle v^2 \rangle / \langle v^2 \rangle_M = 1.0$, is trivial and does not convey any information, as the velocity moments are normalized with Maxwellian moments at the same mean energy.) In contrast with the low density ratio case, at $\eta = 10^{-1}$ [see Fig.~\ref{fig:relax}(b)] all calculated velocity moments are very close to the Maxwellian moments, indicating the development of a nearly Maxwellian $f(v)$ in this case. 

Measurements on the system are carried out only at times when the velocity moments exhibit fluctuations but no drift. Here, the velocity and energy distribution functions (VDF and EDF) of the electrons, $f({\bf v})$ and $f(\varepsilon)$, respectively, are obtained by sampling of the phase space coordinates of individual particles and subsequent averaging over particles and over time. As the electric field is directed along the $x$ axis, $f({\bf v})$ exhibits cylindrical symmetry and reduces to $f(v_x,v_r)$. The energy distribution of electrons is presented in terms of $F(\varepsilon) = f(\varepsilon) / \sqrt{\varepsilon}$. The drift velocity $v_d = \langle v_x \rangle$ and the mean electron energy $\langle \varepsilon \rangle$ can also be obtained from the (phase space) coordinates of individual particles, averaged over particles and time.

\begin{figure}[h!]
\includegraphics[width=1\columnwidth]{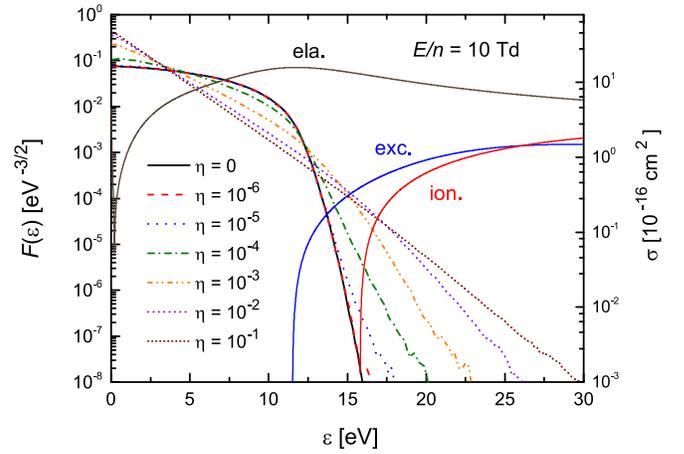}
\caption{\label{fig:edf}
(color online) $F(\varepsilon)$ at $E/n$ = 10 Td, for different values of $\eta$. The additional curves (``ela", ``exc'', and ``ion'') show the cross sections for e$^-$--Ar atom collisions.}
\end{figure}

The energy distribution function, $F(\varepsilon)$, is shown in Fig.~\ref{fig:edf} for different density ratios, $\eta$, at $E/n$ = 10 Td. The data have been obtained by averaging the results of 10 simulation runs each comprising $10^6$ time steps -- this way a ``smooth'' $F(\varepsilon)$ can be generated over about 7 decades of magnitude, at lower values the simulation results become increasingly noisy. The distribution functions obtained at $\eta=0$ and $\eta = 10^{-6}$ are nearly indistinguishable from each other. Compared to the $\eta=0$  case, deviations in the tail of $F(\varepsilon)$ show up at $\eta = 10^{-5}$, while the whole distribution function starts to change its shape at $\eta = 10^{-4}$. With further increasing electron to neutral density ratio the shape changes continuously towards a Maxwellian, represented by a straight line with the given normalization of the EDF. A nearly Maxwellian distribution is reached at $\eta = 10^{-1}$. A similar behavior has been found at the other $E/n$ values considered. The modification of $F(\varepsilon)$ with $\eta$ changes significantly the overlap of the EDF with the cross sections of elementary processes, and thus the collision rate coefficients, $k = \langle \sigma(v) v\rangle$. As a further consequence of this transport coefficients also change remarkably with $\eta$.\cite{Loffhagen1}

\begin{figure}[htb]
\includegraphics[width=0.9\columnwidth]{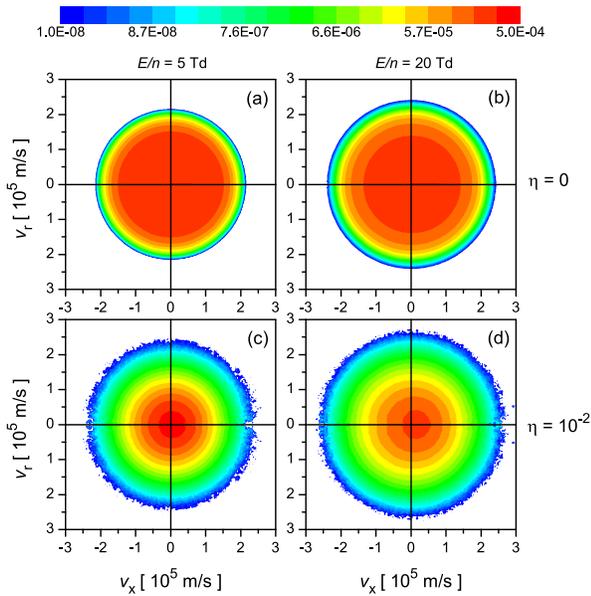}
\caption{\label{fig:vdf}
(color online) $f(v_x,v_r)$ velocity distribution functions of the electrons at $E/n$=5 Td (left column) and 20 Td (right column), at zero (first row) and $\eta = 10^{-2}$ (second row).}
\end{figure}

The changes of the shape of the VDF are illustrated in Fig.~\ref{fig:vdf}. Here we present four cases, characterized by $E/n$ values of 5 Td and 20 Td, and density ratios of $\eta = 0$ and $\eta = 10^{-2}$. The shape of the VDF-s changes notably with the introduction of the electron-electron collisions. Nonetheless, the VDF-s are always nearly isotropic, due to the low applied electric fields. The drift of the distribution (small shift of the VDF towards the positive $x$ direction) is hardly visible in the case of 5 Td, but at 20 Td a clear drift is recognizable. 

\begin{figure}[htb]
\includegraphics[width=1\columnwidth]{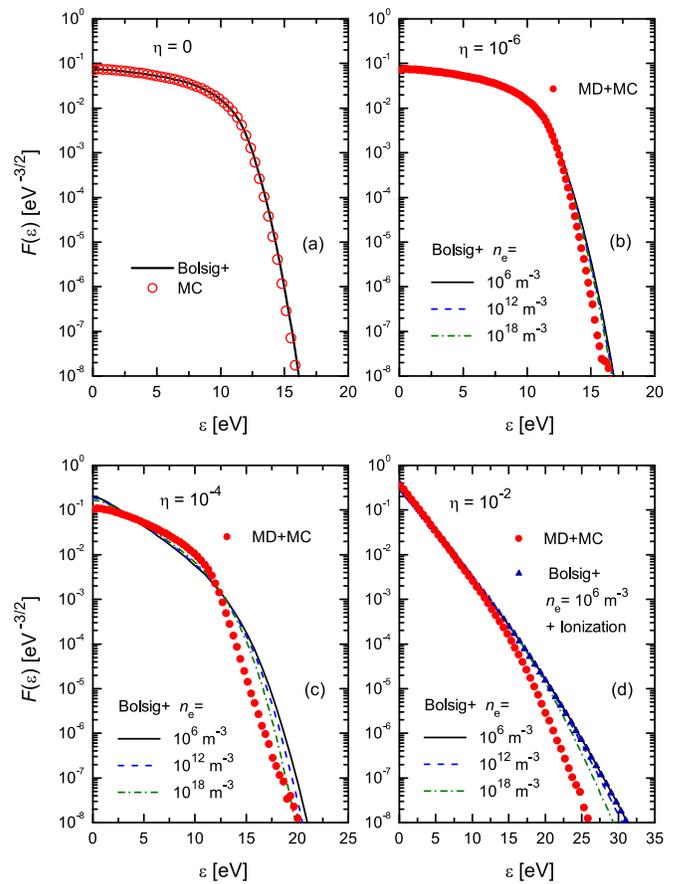}
\caption{\label{fig:mcbe}
(color online) Comparison of the distribution functions obtained with the present method (MD+MC) and from solutions of the Boltzmann equation using the Bolsig+ code \cite{Hagelaar} (with a modified cross section set to treat ionization as a conservative process), for different density ratios: (a) $\eta=0$, (b) $\eta=10^{-6}$, (c) $\eta=10^{-4}$, and (d) $\eta=10^{-2}$. The Bolsig+ calculations are shown for different electron density values indicated in panels (b), (c), and (d). Panel (d) also shows a dataset obtained with the Bolsig+ code with treating ionization as a non-conservative process (triangles). $E/n$ = 10 Td.}
\end{figure}

The energy distribution functions obtained with the present method are compared in Fig.~\ref{fig:mcbe} with solutions of the Boltzmann equation using the Bolsig+ code\cite{Hagelaar}, for $E/n$ = 10 Td. The cross section set of Bolsig+ was modified to treat ionization as a conservative process, just like it is assumed in our particle simulation method. Fig.~\ref{fig:mcbe}(a) shows the results for $\eta=0$, when the particle simulation uses only the MC part. The results obtained via the two approaches are nearly identical in this limit. We note that, when Coulomb effects are considered, Boltzmann solvers assume the presence of screening via space charges, and thus, use the absolute value of the electron density as an input parameter. The results exhibit a weak dependence on the electron density,\cite{Hagelaar}  as it is confirmed as well by the present results displayed in Figs.~\ref{fig:mcbe}(b)-(d). The tail of $F(\varepsilon)$ extends towards higher energies at lower $n_e$ due to the lower degree of screening. Our method does not assume any screening, and thus, one would expect that the results obtained with Bolsig+ method converge towards the present (MD+MC) results in the limit of $n_e \rightarrow 0$. The results, actually, show quite significant deviations from this, which indicate possible issues with the presently available binary collision treatment of Coulomb collisions in MC and BE solutions. The effect of treating the ionization as a non-conservative process has been tested with Bolsig+ at $\eta=10^{-2}$ and $n_e = 10^6$ m$^{-3}$; no observable change of $F(\varepsilon)$ is observed in the data shown in Fig.~\ref{fig:mcbe}(d).  

\begin{figure}[htb]
\includegraphics[width=0.9\columnwidth]{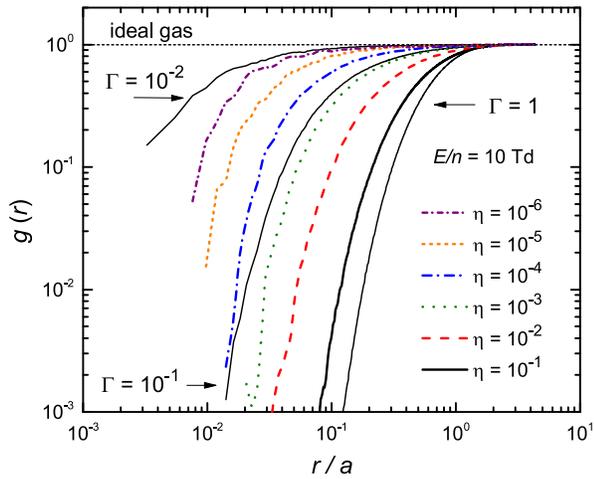}
\caption{\label{fig:pcf}
(color online) (a) PCF-s of the electrons in the swarm at $E/n$=10 Td and different values of $\eta$. The additional curves are results for the one-component (electron) system (without external field and background gas), at the $\Gamma$ values indicated. Distance is normalized by the WS radius $a$.}
\end{figure}

In the $\eta \rightarrow 0$ limit the electron gas behaves like an ideal gas, that is characterized by a pair correlation function $g(r) \equiv 1$, for all $r$. The Coulomb interaction between the electrons creates a ``correlation hole'' [$g(r) < 1$] at small distances, due to the mutual repulsion of the particles. A well-defined correlation hole can already be seen in Fig.~\ref{fig:pcf} at $\eta = 10^{-6}$, despite the fact that $F(\varepsilon)$ very nearly agrees with that at $\eta=0$. With increasing density ratio the correlation hole gradually wides. 

For comparison purposes, in Fig.~\ref{fig:pcf} we also display pair correlations functions calculated for another important model system in plasma physics, the ``one-component plasma'' (OCP) model.\cite{sccs} In the OCP model only one type of the plasma constituents is considered explicitly, the oppositely charged particles are assumed to form an unpolarizable, neutralizing background. The PCF-s are shown for different values of the coupling parameter 
\begin{equation}
\Gamma= e^2 / (4 \pi \epsilon_0 a k_B T),
\end{equation} 
where $a =  (3/4 \pi n_e)^{1/3}$ is the Wigner-Seitz (WS) radius, $k_B$ is the Boltzmann constant, and $T$ is the temperature. The PCF-s calculated for the electron swarms, for $10^{-6} \leq \eta \leq 10^{-1}$ are bound by the PCF-s characterizing the OCP at $\Gamma=0.01$ and $\Gamma$=1 -- a range of $\Gamma$ that belongs to the {\it non-ideal plasma} domain.

\begin{figure}[htb]
\includegraphics[width=0.9\columnwidth]{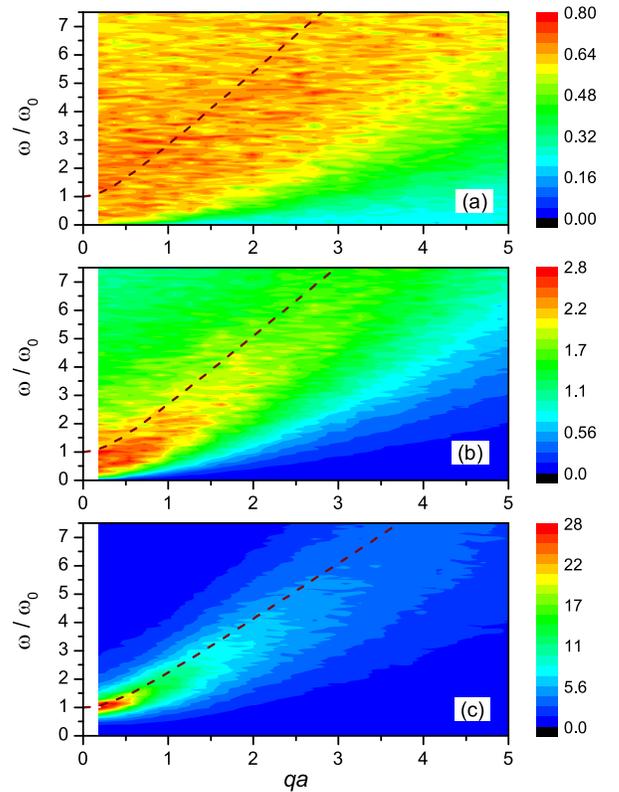}
\caption{\label{fig:skw}
(color online) Spectra of longitudinal current fluctuations, $L(q,\omega)$, at $E/n$ = 10 Td and $\eta = 10^{-5}$ (a), $10^{-4}$ (b), and $10^{-2}$ (c). The dashed lines represent the Bohm-Gross dispersion relation, with corresponding values of $\langle \varepsilon \rangle$. The spectra are given in arbitrary units, the color scale is linear, and the wave number is normalized by the WS radius $a$.}
\end{figure}

The MD approach makes it possible as well to study the emergence of waves (plasma oscillations) within the swarm, via computation of the density and current fluctuation spectra. We calculate the spectra of longitudinal current fluctuations, $L(q,\omega)$, for a discrete set of wave numbers $q = p (2 \pi / H) = p q_{min},~p=1,2,...$ , accommodated by the simulation box of edge length $H$. To accomplish this calculation we collect data during each time step of the simulation for the microscopic current 
\begin{equation}
\lambda(q,t)= \sum_j v_{j x}(t) {\rm e}^ { i q x_j(t)},
\end{equation}
where $x_j$ and $v_j$ are the position and velocity of the $j$-th particle. These data sequences are subsequently Fourier analyzed:  \cite{Hansen,review}
\begin{equation}\label{eq:sp1}
L(q,\omega) = \frac{1}{2 \pi N} \lim_{\tau \rightarrow \infty}
\frac{1}{\tau} | \lambda(q,\omega) |^2,
\end{equation}
where $\tau$ is the length of data recording period and $\lambda(q,\omega) = {\cal{F}} \bigl[ \lambda(q,t) \bigr]$. Collective excitations (waves) appear as peaks in $\lambda(q,\omega)$. The fluctuation spectra obtained at $E/n$ = 10 Td, for $\eta = 10^{-5}$, $10^{-4}$, and $10^{-2}$ are displayed in Fig.~\ref{fig:skw}(a,b,c) in the form of a color maps. At low $\eta$ the energy is spread widely in the ($q,\omega$) plane, but with increasing electron to neutral density ratio we can follow the development of a pronounced collective mode (fully developed at $\eta=10^{-2}$). At $q \rightarrow 0$ the mode frequency equals the plasma frequency, $\omega = \omega_0$, with increasing wave number $\omega$ increases, following closely the Bohm-Gross dispersion relation of {\it warm electrostatic waves}:
\begin{equation}
\omega^2 = \omega_0^2 + \frac{3 k_B T_e}{m_e} q^2.
\end{equation}
This dispersion relation, calculated with the mean electron energy values: $\langle \varepsilon \rangle = \frac{3}{2} k_B T_e \cong$  5.25 eV for $\eta = 10^{-2}$, 4.65 eV for $\eta = 10^{-4}$, and 2.98 eV for $\eta = 10^{-2}$ are superimposed as dashed lines on the color maps of $L(q,\omega)$ in Fig.~\ref{fig:skw}. 

Finally we note that the new method is computationally much more expensive than efficient Boltzmann solvers and run times typically exceed by one order of magnitude even the run times of Monte Carlo codes, as most of the simulation time is devoted to handling of many-particle effects. For the conditions and system parameters studied here, a run with $10^6$ time steps uses about two days of CPU time. These runtimes, however, do not represent an issue when accuracy has a priority over fast computations based on approximate schemes.

\section{Summary}

In summary, a {\it new computational method}, from the combination of the Molecular Dynamics and Monte Carlo techniques, has been proposed to describe electron swarms, in the presence of appreciable electron-electron interaction. The method is based on first principles and provides a fully kinetic description of the system without the need of introducing approximations in the treatment of Coulomb collisions. 

The capabilities of the method have been illustrated via calculations of electron swarm characteristics in argon gas, at low reduced electric fields. The simulations made it possible to follow the modifications of the velocity and energy distribution functions of the electrons across the different regimes of the electron to neutral density ratio. Besides calculating the ``usual'' swarm characteristics, the method also allowed (i) to study the development of correlations in the electron gas with increasing $\eta$, as indicated by the pair correlation function, and (ii) identification of a developing collective mode at significant electron to neutral density ratios. 

The new method can be extended to include the temporal growth of the electron density, as well as to describe swarm behavior in high-frequency fields. Incorporation of the screening by the plasma is straightforward, via changing the Coulomb interaction potential to Yukawa type in the MD part of the code. 

The differences of the energy distribution functions obtained by the present method and via Bolsig+ point to the possible issues with the conventional binary approach for Coulomb collisions in Monte Carlo simulations and in Boltzmann equation solutions. Detailed investigation of these issues is planned as a future work.

\acknowledgements

This work has been supported by the Grant OTKA K-105467.

\end{document}